# "The Intangible Victory", Interactive Audiovisual Installation

**K. Tsioutas**
Computer Science Department, Athens University of Economics and Business, Athens, Greece
E-mail: ktsioutas@aueb.gr, Tel +30 6934525150

**P. Pangalos**
Interior Design Department, University of West Attika, Athens, Greece
E-mail: pa.pa@uniwa.gr

**K. Tiligadis**
Audio and Visual Arts Department, Ionian University, Corfu, Greece
E-mail: gustil@ionio.gr

**A. Sitorengo**
Interior Design Department, University of West Attika, Athens, Greece
E-mail: asitor@uniwa.gr

**Abstract**

"Intangible Victory" is an audiovisual installation in the form of the intangible being of the "Victory of Samothrace" that uses interactive digital media. Specifically, through this installation, we redefine the visual symbolism of the ancient sculpture, paying attention to time as a wear factor (entropy) and the special importance of the void as an absence of the sculptural form. Emptiness completes the intangible essence of the sculpture in the field of symbolism as well as in that of artistic significance for the interpretation of the work today. The function of the void and the interaction of the viewer with the work, causes the emergence of a new experience-dialogue between space and time. The use of digital media and technology reveals the absence of the sculptural form as it is visualized in the "Victory of Samothrace". The sculptural form is reconstructed from fibers in space in a cylindrical arrangement. The form is rendered with colored strings - conductive sensors, that allow the visitor to interact with the work, creating a sound environment through movement. The sound completely replaces the volume, as the void of the sculptural form together with the viewer in unison present an audiovisual symbolism of the "Victory of Samothrace".

## 1. INTRODUCTION

A large amount of archaeological Greek masterpieces is held in various European museums exhibiting the greatness of ancient Greek Civilization. The statue of "The Winged Victory of Samothrace" which is held in the museum of Louvre, is one of the most famous statues from ancient Greece **[1] [2]**. For the needs of the periodic exhibition Symbols and Iconic Ruins **[3]**, held in the Greek National Museum of Modern Art, we designed the "The Intangible Victory" an Interactive Audio Installation. The goal of our approach was to reproduce the atmosphere and the materiality, the ancient statue reveals, through the timeline from the ancient years to today. The "The Intangible Victory" was shaped by multiple white threads, representing the mantle of the statue in an intangible state. The form shaped by the threads, was designed on a base formed like a boat as the original statue was. The visitors could touch the threads and interact with the sculpture to integrate the artistic sensation of it. As described by multiple visitors of the exhibition "the sound heard while touching the threads, takes us to a mythology of this deity but in a different way, not in the traditional, in a more emotional, visual and auditory way". The title of Intangible Victory comes from the concept of victory which is many time pointless (intangible) as it does not matter if you win or lose.

The statue, made of marble from Paros (one of the finest marbles of Greece), represents the goddess of Victory about to alight on a ship whose sailors have just won a sea battle. It probably dates to 190 BC and was commissioned to celebrate that victory. The winners, perhaps the inhabitants of the island of Rhodes, erected it in Samothrace to thank the Great Gods of the island, the Cabeiri, who were worshipped throughout the Greek world.

It was discovered in 1863 by Charles Champoiseau, broken and incomplete. The head and the arms were missing and were never found. The statue is breathtaking while it exposes the greatness of ancient Greek art. Although as a masterpiece, the winged victory leaves no space for any attempt of reproduction and redesign, it can be a great motive for inspiration. As many ancient Greek treasures have been discovered with many missing or broken parts, strongly affected by time, wars and all kinds of disasters, the Winged Victory is a ruined entity, taken away from its motherland, missing from the place it was originally found.

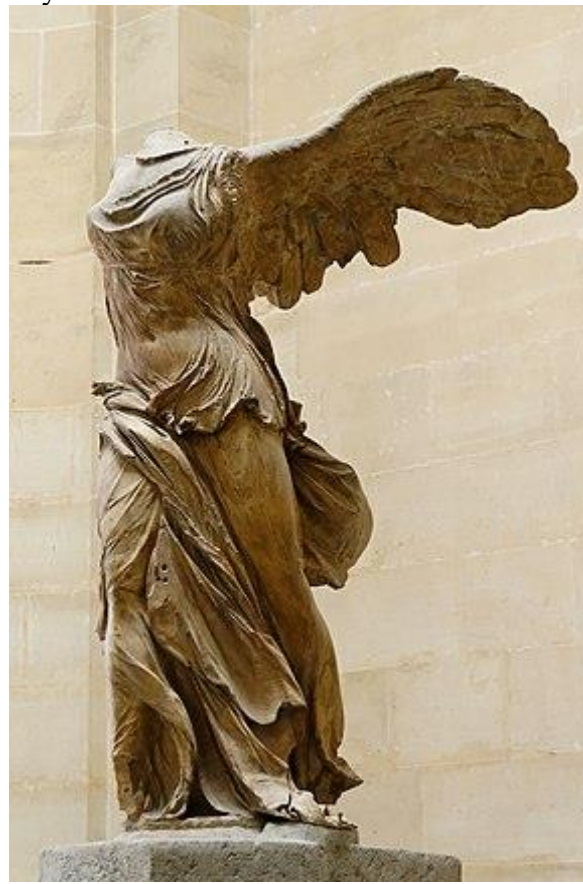

Figure 1: The Winged Victory Statue in Louvre Museum.

## 2. THE RUINS EXHIBITON

In the history of civilisation, works of art and architecture have been created or simply used as symbols of religious, political or economic power. When a form becomes a means of reference of a symbol, then it may reach such a level of transcendental abolition of its materiality that it can be formulated even with abstract archetypal geometry, as is the case with mathematical symbols. In this regard, the mediating role of art is often crucial in terms of the representation of symbolic structures, which - among other things - contribute to the consolidation of a balance that gives chaos a sense of order and freedom.

The ruins of each era, made of marble, stone, brick, metal, or concrete, are spatial manifestations of worn matter with significant symbolic added value. Indeed, the symbolic reading of the ruins has been largely imbued with the historical and philosophical layers of numerous ideologies. The ruin not only tells a story but is mainly used in the production of the necessary story to be told. As a tool for managing memory and a means of disseminating selected information of yesterday, it is the par excellence case of regular and strategic influence of collective memory and consciousness. National ideologies often resort to ruins to prove that they have long before acquired self-existence and to strengthen the symbolic and ideological ties of the community. Even when this is not possible, reality can be altered and replaced by an illusion of reality. Tradition may be invented and historical continuity can be established in a fictitious relationship with an artificial, idealised past, which always contains the ideals of today as goals to be conquered. A great example is the rock of the Acropolis of Athens, which possesses multiple historical identities, but is mainly associated with the values of the ancient Greek republic and is arbitrarily called "Sacred", as a symbolic work of classic art, a floating imaginary city, which could, not be based on specific territory.

This idealisation is repeated in the case of the Victory of Samothrace. It is a Victory that today distances itself from Samothrace, to refer to the universal idea of victory, as Picasso's Guernica refers to the flame of every war. Correspondingly, the Intangible Victory of Heimer & Alz (Andreas Sitorengo, Kostantinos Tiligadis), through the distance from materiality and mainly from the property of matter to fill the space, to fill the void, to satisfy the desires, to generate glory, creates the conditions to understand that any conflict, apart from the victors [winners], leaves behind defeated ones [losers]. This intangible victory may be a good reason to seek a different victory, which will not produce pain in its passage, but will cure it.

## 3. THE SCULPTURE
We chose to use plexiglass for shaping the basic structure of the sculpture which represents the intangible form of the Winged Victory statue while the threads represent the mantle the goddess wears. Three parts of plexiglass, the base, the top and the column are connected together and multiple threads are passed from the base to the top one by one as shown in the pictures. The base is shaped as a ship as the real statue is placed on. The sculpture is integrated through the transition to the digital age and the use of technology, sensors programming, physical computing and sound design. All these elements, form the structure of the message that we want to transmit to the visitor so she/he can get inside this symbolic cluster of art, technology and science and interact with it. The sculpture in its wholeness, is described by its form through the artistic language, and it is complemented by the digital narration of the media, thus the programming language and sound compositions.

## 4. PROGRAMMING AND PHYSICAL COMPUTING
One of the basic design goals of the sculpture was its interactive character. The sculpture should interact with the audience and a dynamically modulated relationship should be established in time and space. For the interaction, conductive threads were used between the white threads. The conductive threads are used to represent the physical instance of the statue as it is placed in the stern of a ship, air is blowing and moves randomly the mantle and sea waves are crashing on the ship. Conductive threads are used in many interactive projects where textile is used for interaction. These threads allow the electrical current to pass through and can easily be used as touch sensors connected in electrical circuits. They have an extremely low ohm resistance and can easily be connected to microcontrollers like the Arduino Platform **[5]**. The threads were about 8 m each and they were passed three times from the bottom to the top of the development. The visitors were urged to touch the threads which while touched, triggered four different audio compositions representing, the sound of the wind, the sound of the sea, the sound of the mantle forced by the wind and the sound of the marble hit by strong wind. The sounds were composed by Nikolaos Sidiropoulos, pianist and music composer. The threads could act like antennas sensing the existence of the visitors while they were close to the sculpture. Through physical computing the Arduino platform was programmed to receive signals from the threads and process them in order to trigger the composed sounds. Additionally real time algorithms were running over MaxMsp **[6]** programming interface to produce sums of sinusoidal signals with low audio frequencies which produced an evocative sound environment.

## 5. INTERACTION WITH THE AUDIENCE

The sculpture was exhibited for six months in the Greek National Museum of Modern Arts. Hundreds of people visited the museum during the exhibition and had the chance to interact with it. People were touching the strings in multiple ways, some times they were just softly touching them with their fingers and other times they were pulling them like they were strings of an harp. As many visitors mentioned, the sculpture was revealing an abstract sense of a musical instrument. Many times, the strings were broken because of the strength the visitors used to pull them. As Nefeli Katsarou co – artist of the exhibition commented, "when one touches them, an eerie music is heard, which, takes us to a mythology of this deity, but in a different way, not with traditional, in a more emotional, visual and auditory way "**[4]**.

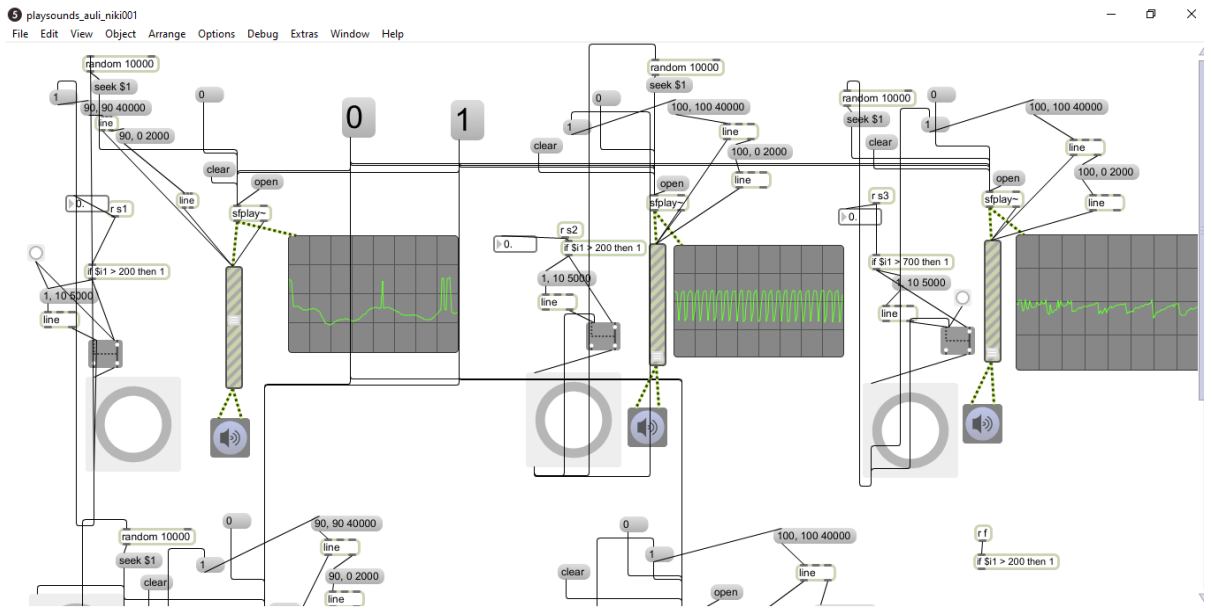

**Figure 2: The MaxMsp programming Environment**

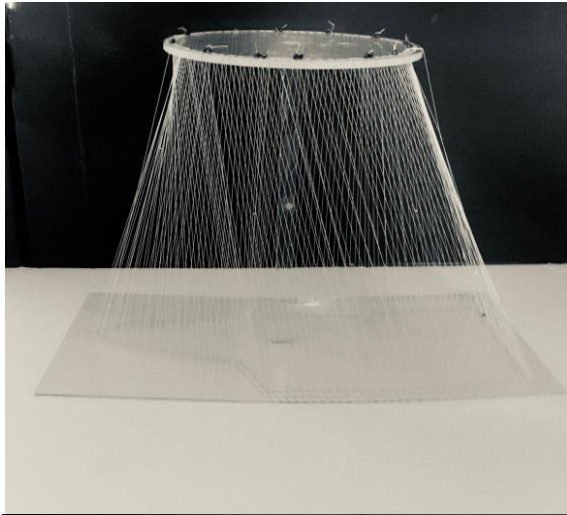
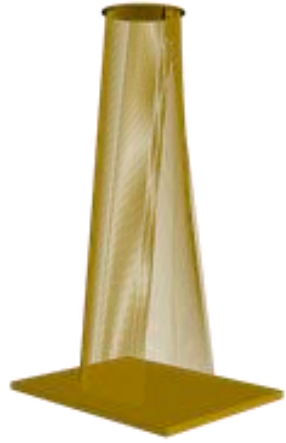
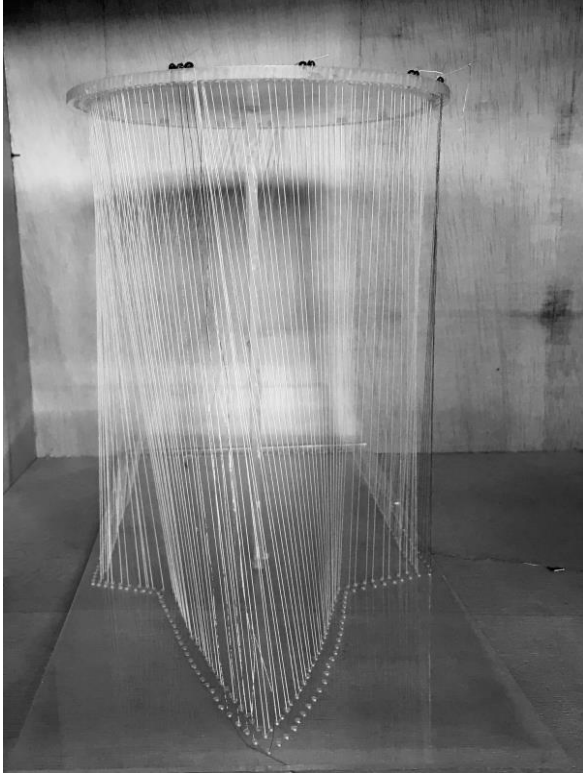
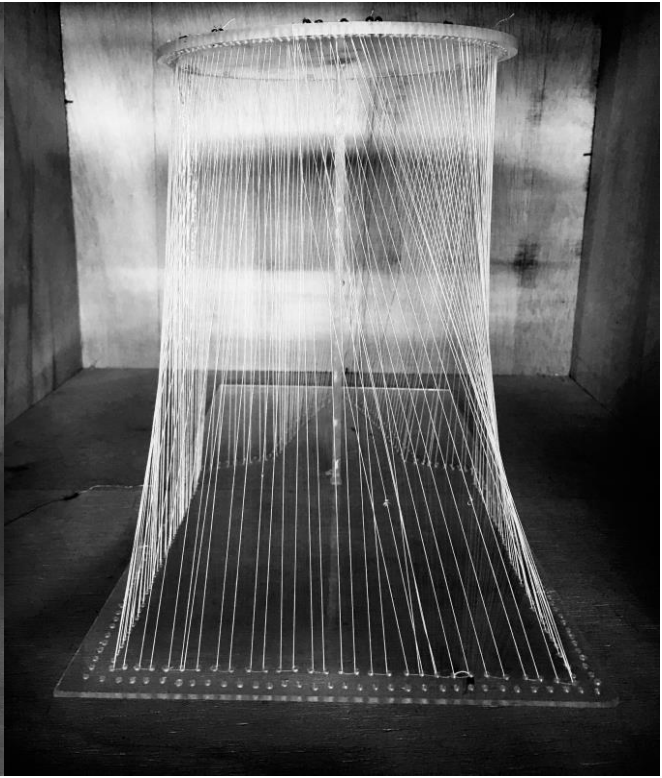

**Figure 3: Drafts of the sculpture**

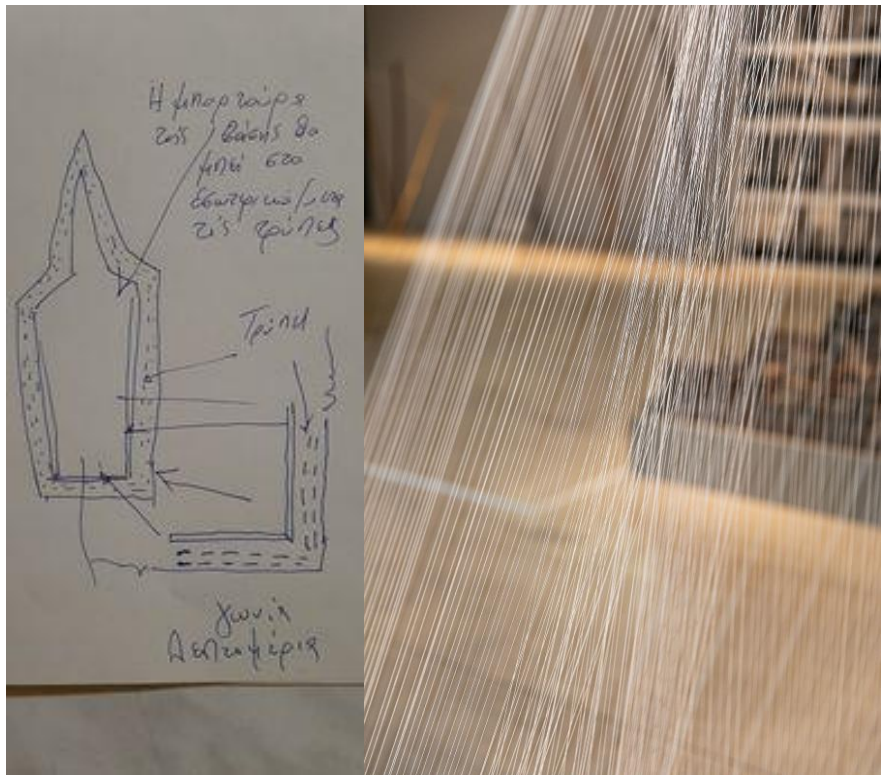
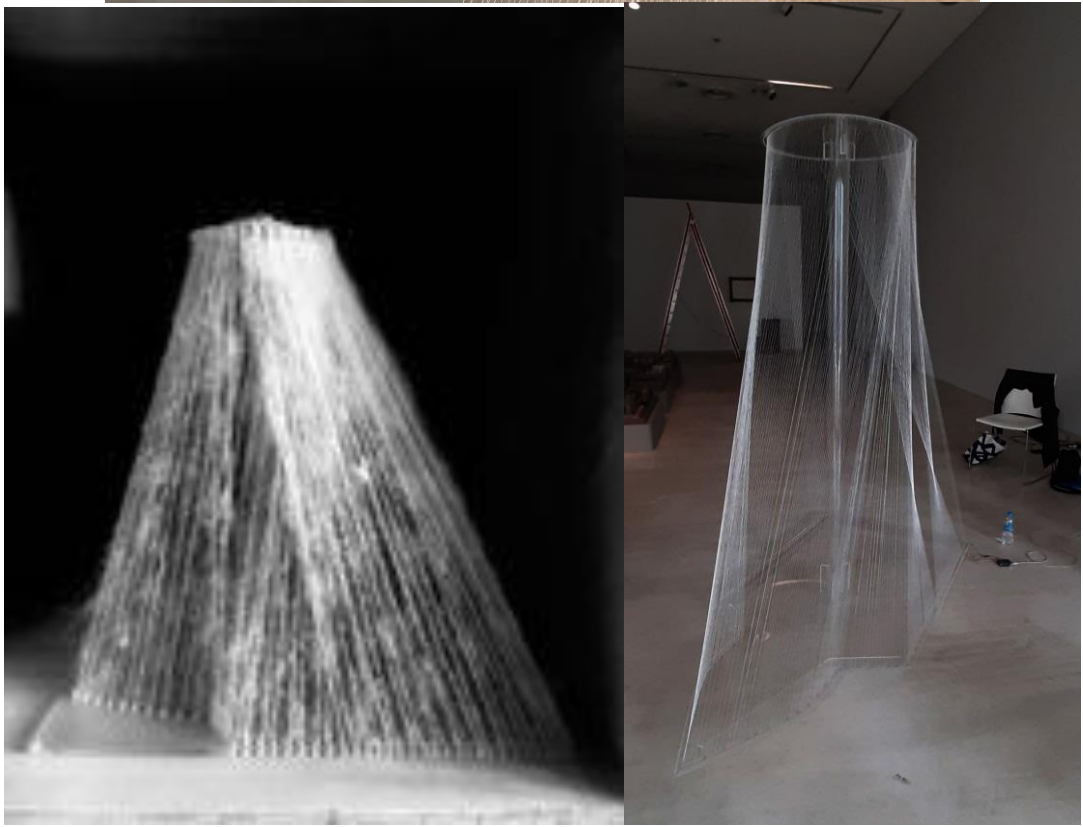

**Figure 4: Multiple experimental forms**

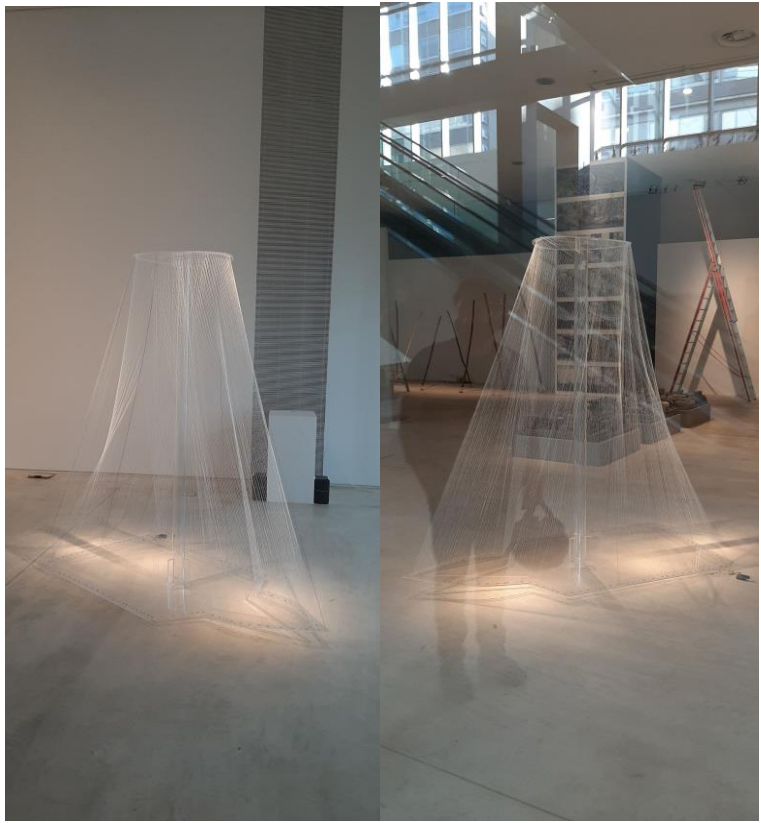

## 6. CONCLUSION

In this project, we propose an audiovisual installation in the form of the intangible being of the Victory of Samothrace by using interactive digital media. Specifically, in the interactive installation we redefine the visual symbolism of the ancient sculpture, paying attention to time as a wear factor (entropy) and the special importance of the void as an absence of the sculptural form. The finding of emptiness completes the intangible essence of the sculpture in the field of symbolism but also of the artistic significance for the interpretation of the work today. The function of the void and the interaction of the viewer with the work causes the emergence of a new experience-dialogue between space and time. The use of digital media and technology reveals the absence of the sculptural form as it is visualized in the "Victory of Samothrace". The sculptural form is reconstructed from fibres in the space in a cylindrical arrangement. The form will be rendered with coloured strings - conductive sensors, that will allow the visitor to interact with the work creating a sound environment through its movement. The sound will completely replace the volume as the void of the sculptural form together with the viewer as in a composition will present an audiovisual symbolism of the "Victory of Samothrace".